\documentclass[12pt]{amsart}
\usepackage{amssymb}
\usepackage{amsmath,amsthm,amsfonts}
\usepackage[applemac]{inputenc}
\usepackage{graphicx}
\usepackage{bbm}
\usepackage{pdfsync}
\usepackage{fancyhdr}
\usepackage{mathrsfs}
\usepackage{slashed}
\usepackage{color}
\usepackage{xcolor}
\usepackage[breaklinks,colorlinks,backref]{hyperref}

\hypersetup{
colorlinks, 
linktoc=all, 
linkcolor=blue, 
citecolor=hyptxt,
urlcolor=blue}
\hypersetup{
citebordercolor=,
filebordercolor=green,
linkbordercolor=blue
}
\definecolor{hyptxt}{rgb}{0.7, 0.4, 0.9}

\usepackage{bibtopic}

\usepackage[full]{textcomp} 
\usepackage{fbb} 
\usepackage[varqu,varl]{inconsolata}
\usepackage[libertine,bigdelims,vvarbb]{newtxmath}
\usepackage[cal=boondoxo]{mathalfa}
\usepackage[T1]{fontenc}
\useosf

\newtheorem{prop}{Proposition}[section]

\newcommand{\beprop}{\begin{prop}}
\newcommand{\enprop}{\end{prop}}
\newcommand{\bprf}{\begin{proof}}
\newcommand{\eprf}{\end{proof}}

\newcommand{\ket}[1]{|\kern.3ex#1\kern.3ex\rangle}
\newcommand{\bra}[1]{\langle\kern.3ex #1 \kern.3ex|}
\newcommand{\scalar}[2]{\langle\kern.3ex #1 \kern.3ex|\kern.3ex#2\kern.3ex\rangle}

\definecolor{hervecolor}{rgb}{0.8,0,0.7}

\numberwithin{equation}{section}

\def\1{\mbox{I\hspace{-.15em}1}}

\def\K{\mathcal{K}}

\def\K{\mathcal{K}}
\def\b{\begin{equation}}
\def\e{\end{equation}}

\begin{document}
\date{\today}
\title{Polarization tensor in de Sitter gauge gravity}
\author{R. Raziani, M.V. Takook}

\address{\emph{ APC, UMR 7164}\\
\emph{Universit\'e de Paris} \\
\emph{75205 Paris, France}}

\email{ r.raziani@razi.ac.ir, takook@apc.in2p3.fr}

{\abstract{
The gauge theory of the de Sitter group, SO$(1,4)$, in the ambient space formalism has been considered in this article. This method is essential to constructing the de Sitter super-conformal gravity and Quantum gravity. $10$ gauge vector fields are needed, corresponding to $10$ generators of the de Sitter group. Using the gauge-invariant Lagrangian, the field equation of these vector fields has been obtained. The gauge vector field solutions are recalled. Then, the spin-$2$ gauge potentials are constructed from the gauge vector field. There are two possibilities for presenting this tensor field: rank-$2$ symmetric and mixed symmetry rank-$3$ tensor fields. To preserve the conformal transformation, a spin-$2$ field must be represented by a mixed symmetry rank-$3$ tensor field, $\mathcal {K}_{\alpha\beta\gamma}$. This tensor field has been rewritten using a generalized polarization tensor field and a de Sitter plane wave. This generalized polarization tensor field has been calculated as a combination of vector polarization, $\mathcal {E}_{\alpha}$, and tensor polarization of rank-$2$, $\mathcal {E}_{\alpha\beta}$, which can be used in the gravitational wave consideration. For the construction of this polarization tensor, the arbitrary constant vector fields appear. We fix it so that, in the limit, $H=0$, one obtains the polarization tensor in Minkowski space-time. It has been shown that under some simple conditions, the spin-$2$ mixed symmetry rank-$3$ tensor field can be simultaneously transformed by the unitary irreducible representation of de Sitter and conformal groups, SO$(2,4)$.}}

\maketitle

{\it Proposed PACS numbers}: 04.62.+v, 12.60.Jv, 11.10.Cd, 12.10.-g

\tableofcontents

\section{Introduction}

Astrophysical data shows that our universe is expanding with a positive acceleration \cite{acce}. Therefore, the universe can be described through the de Sitter (dS) space-time in the first approximation \cite{perme,hebb,hem,berna}. On the other hand, the recent observational data by BICEP2 confirm that in the evolution of the early universe, a dS metric appeared \cite{parde}. Thus, it is crucial to formulate the quantum field theory (QFT) in the dS space. The QFT and linear quantum gravity in dS ambient space formalism have been presented elsewhere \cite{ta1403,ta2020}.

The massless fields in Minkowski space-time are propagating on the light cone. So, their field equations are invariant under the conformal transformations. For spin, $s\geq 1$, they are also invariant under the gauge transformations formulated by the Yang-Mills gauge theory. In dS space, mass is not a constant parameter for the set of observable transformations under the dS group, SO$(1,4)$. It is important to note that although the conformal invariance guarantees the light-cone propagation, the inverse is not always true \cite{fr84}. However, the concept of light-cone propagation does exist in dS space-time.

In linear conformal gravity, the propagating modes must be a pair of massless particles with helicities $\pm 2$. The spin-$2$ gauge fields have been studied in \cite{SRI84}. A natural choice for such a field is a symmetric tensor field of rank-$2$. Barut and B{\"o}hm \cite{babo} have proved that the value of the conformal Casimir operator for the physical representation of the conformal group is $9$. Based on the work of Binegar et al. \cite{bifrhe}, however, for any tensor field of rank-$2$, this value becomes $8$. Therefore, one of the best forms of gravitational field corresponding to the physical representations of the conformal group is a mixed symmetry rank-$3$ tensor field with conformal degree zero \cite{bifrhe}.

To construct the super-gravity, it is required to formulate the linear gravitational field theory similar to other gauge theories, {\it e.g.} Yang-Mills gauge theory. In this regard, Nieto presented the gauge theory of the dS group in the intrinsic coordinate system \cite{nieto}. This article introduces the dS group gauge theory in the ambient space formalism, which can also be conformal invariant. In the ambient space formalism, the transverse gauge-covariant derivative and the gauge-invariant Lagrangian density can be defined in line with what is proposed before \cite{ta1403}. The transverse gauge-covariant derivative contains two parts: the first part leads to a constant space-time curvature (dS background), and the second leads to the gauge potentials propagating on the dS space-time.

$10$ gauge potentials vector fields exist, corresponding to the $10$ generators of the dS group. The linear gravity or spin-$2$ field can be constructed from these gauge potentials, similar to what was reported in \cite{mama}. There are two possibilities for constructed spin-$2$ gauge potentials: the rank-$2$ symmetric tensor field, $H_{\alpha\beta}$, and the mixed symmetry rank-$3$ tensor field, $ \mathcal{K}_{\alpha\beta\gamma}$. The conformal invariant is only preserved in the latter. Here, we present the construction of these fields.

Section \ref{nota} reviews the notation of the dS ambient space formalism. The gauge-covariant derivative and the gauge-invariant action are presented in Section \ref{III}. In Section \ref{IV}, the gauge vector fields equation has been obtained using the Euler-Lagrange equation. At the end of this Section, three types of fields are derived: gauge vector field, gauge rank-$2$ tensor field, and gauge rank-$3$ tensor field. Finally, the results are discussed in the conclusion part. Some practical mathematical details of our calculations are given in the appendices.

\section{de Sitter ambient space formalism} \label{nota}

The dS space-time is a maximally symmetric solution of the vacuum Einstein equations with a positive cosmological constant $\Lambda$. It can be viewed as a one-sheeted hyperboloid embedded in a five-dimensional Minkowski space-time:
\begin{equation} \label{dSs}
M_H=\left\lbrace x \in \mathrm{I\hspace{-.15em}R}^5 |\; \; x \cdot
x=\eta_{\alpha\beta} x^\alpha x^\beta =-H^{-2}\right\rbrace,
\end{equation}
where, $\eta_{\alpha\beta}=$diag$(1,-1,-1,-1,-1)$, $\alpha,\beta=0,1,2,3,4,$
and $H$ is the Hubble constant parameter. Hereafter, for simplicity, we set $H=1$. The dS metric is
\begin{equation}
ds^2=\eta_{\alpha\beta}dx^{\alpha}dx^{\beta}|_{x\cdot x=-1}=g_{\mu
\nu}^{dS}dX^{\mu}dX^{\nu},\;\; \mu=0,1,2,3,
\end{equation}
where $x^\alpha$ stands for ambient space formalism, and $X^\mu$ is used
for dS intrinsic coordinates. We use the ambient space formalism for ease of manipulation, which is very similar to Minkowski space. The main points of this formalism can be listed as follows:
\begin{itemize}
\item (A) The projection operator from $5$-dimensional Minkowski space to $4$-dimensional dS hyperboloid is defined as:
\begin{equation}
\theta
_{\alpha\beta}=\eta_{\alpha\beta}+x_\alpha x_\beta.
\end{equation}
\item (B) The transverse-covariant derivative acting on a tensor field of rank-$l$ is defined by:
\begin{equation}\label{dscdrt}
\nabla^\top_\beta T_{\alpha_1....\alpha_l}\equiv \partial^\top_\beta
T_{\alpha_1....\alpha_l}-\sum_{n=1}^{l}x_{\alpha_n}T_{\alpha_1..%
\alpha_{n-1}\beta\alpha_{n+1}..\alpha_l},
\end{equation}
where $\partial^\top_\beta=\theta_\beta^{\;\;\alpha}\partial_\alpha=\partial_\beta+H^{2}x_\alpha x\cdot\partial\,,\;\; x\cdot\partial^\top=0$.
\item (C) It is worth to mention that: $x\cdot\nabla^\top=0$.
\end{itemize}

The dS kinematical group is the $10$-parameters group SO$(1,4)$. There are two Casimir operators for this group \cite{tak,tho}, the second-order Casimir operator is:
\begin{equation}\label{casi1}
Q^{(1)}=-\frac{1}{2}L_{\alpha\beta}L^{\alpha\beta},\;\; \alpha,
\beta=0,1,2,3,4,
\end{equation}
and the fourth-order Casimir operator is:
\begin{equation}
Q^{(2)}=-W_\alpha W^\alpha\;\;,\;\;W_\alpha =\frac{1}{8} \epsilon_{\alpha%
\beta\gamma\delta\eta} L^{\beta\gamma}L^{\delta\eta},
\end{equation}
where $\epsilon_{\alpha\beta\gamma\delta\eta}$ is the antisymmetric tensor in $5$-dimensional Minkowski space.
$L_{\alpha\beta}$ are the
infinitesimal generators of dS group, $L_{\alpha\beta}=M_{\alpha\beta}+S_{\alpha\beta}$. The orbital part $M_{\alpha\beta}$, is defined as follows:
\begin{equation}\label{genm}
M_{\alpha \beta}=-i(x_\alpha \partial_\beta-x_\beta
\partial_\alpha)=-i(x_\alpha\partial^\top_\beta-x_\beta
\partial^\top_\alpha).
\end{equation}
The spinorial
part, $S_{\alpha\beta}$, with an integer spin $s=l=0,1,\cdots$, is read as \cite{frhe,gaha}:
\begin{equation}\label{gens}
S_{\alpha \beta}^{(l)}\mathcal{K}_{\gamma_1......\gamma_l}=-i\sum^l_{i=1}
\left(\eta_{\alpha\gamma_i} \mathcal{K}_{\gamma_1....(\gamma_i\rightarrow%
\beta).... \gamma_l}-\eta_{\beta\gamma_i} \mathcal{K}_{\gamma_1....(\gamma_i%
\rightarrow \alpha).... \gamma_l}\right),
\end{equation}
where $(\gamma_i\rightarrow\beta)$ means $\gamma_i$ is replaced
by $\beta$.

\setcounter{equation}{0}
\section{Gauge-invariant action}\label{III}

The local or gauge symmetry is a fundamental basis for explaining the electromagnetic, weak, and strong nuclear interactions by the gauge vector fields in Minkowski space-time. Here, we construct the gauge theory in dS ambient space formalism for the dS group SO$(1,4)$.
The notation is changed for simplicity, and the dS group generators are defined as follows:
\begin{equation*}
L_{\alpha\beta}=M_{\alpha\beta}+S_{\alpha\beta}\equiv X_A, \;\; A=1,2,...,10.
\end{equation*}
In this case, the commutation relation can be reformulated as:
\begin{equation*}
[X_A,X_B]=f_{BA}^{\;\;\;\;\;C}X_C,
\end{equation*}
where $f_{BA}^{\;\;\;\;\;C}$ is the Lie algebra structure constant, which is explicitly presented in \cite{nieto}. We have $10$ gauge vector fields $K_\alpha^{\;\;A}$. As these gauge vector fields exist on the dS
hyperboloid, they should be transverse: $x\cdot K^A \equiv x^\alpha K_\alpha^{\;\;A}=0$.
From the gauge theory point of view, the transverse gauge-covariant derivative in the ambient space notation can be defined as:
\begin{equation}\label{gcd2}
D^{K}_\alpha=\nabla^\top_\alpha+i K_\alpha^{\;\;A} X_A,
\end{equation}
where $\nabla^\top_\alpha$ is defined in equation (\ref{dscdrt}).
There exist $40$ degrees of freedom for these gauge vector
fields. So, they can be recognized as the connection coefficients in the general
relativity framework. The construction of the gauge theory in its canonical manner can be reproduced. Under a local infinitesimal gauge transformation generated by $\epsilon^A(x)X_A$, one can obtain the following gauge transformation:
\begin{equation} \label{gatra3mif}
\delta_{\epsilon} {K}_\alpha^{\;\;A}=D^{K}
_\alpha\epsilon^A=\nabla^\top_\alpha \epsilon^A+ f_{CB}^{\;\;\;\;A} {
K}_\alpha^{\;\;C}\epsilon^B.
\end{equation}
The curvature $\mathcal{R}$, with values in the Lie algebra of the dS group,
is defined by:
\begin{equation*}
\mathcal{R}(D^{K}_\alpha,D^{K}_\beta)=-[D^{K}
_\alpha,D^{K}_\beta]=R_{\alpha\beta}^{\;\;\;\; A}X_A,
\end{equation*}
where
\begin{equation} \label{cuso14}
R_{\alpha\beta}^{\;\;\;\; A}=\nabla^\top_\alpha K
_\beta^{\;\;A}-\nabla^\top_\beta K_\alpha^{\;\;A}+ K
_\beta^{\;\;B}K_\alpha^{\;\;C}f_{BC}^{\;\;\;\;A}.
\end{equation}
In the ambient space formalism, the gauge-invariant action or Lagrangian density for this curvature can be defined as \cite{mama,van,frts}:
\b \label{gias2r31gg} S[K]=\int d\mu(x)\left(R_{\alpha\beta}^{\;\;\;\;\;A}q_{AB}R^{\alpha\beta B}\right),\e
where $d\mu(x)$ is dS invariant volume element and $q_{AB}$ is a numerical constant.
\setcounter{equation}{0}
\section{ Gauge fields} \label{IV}

Using the Euler-Lagrange equation for (\ref{gias2r31gg}), the field equation can be obtained as (see appendix A):
\b \label{casira2}-Q_0 K_{\alpha}^{\;\;A}+ 2K_{ \alpha }^{\;\;A}-x_{\alpha} \partial^\top \cdot K^{\;A} - \partial^\top_\alpha\partial^\top \cdot K^{\;A}=0 ,\e
where $$Q_0=-\frac{1
}{2}M_{\alpha\beta}M^{\alpha\beta}=-\partial^\top\cdot \partial^\top.$$
If these vector fields are desired to propagate on the dS hyperboloid and transform as a UIR of the dS group, they must satisfy the transversality and the divergencelessness conditions \cite{ta1403},
$$x\cdot K^{\;A}=0,\;\;\partial^\top \cdot K^{\;A}=0,$$
then we have:
\b \label{fer31} (Q_0-2) K_{\alpha}^{\;\;A}=0.\e

The four generators $L_{4\mu}, (\mu=0,1,2,3)$ become the translation generators $P^\mu$ in the flat space-time limit, and the gauge fields $ E_\alpha^\mu$, which correspond to them, play the role of the linear gravitational wave. The six generators $L_{\mu\nu}$ in the null curvature limit are related to the Lorentz group generators, SO$(1,3)$. The "rotational" gauge fields $W_\alpha^{\mu\nu}$ correspond to the generators $L_{\mu\nu}$.With these fields, the gauge-covariant derivative (\ref{gcd2}) can be written in the following form:
\begin{equation}
D^{K}_\alpha=\nabla^\top_\alpha+i {K}_\alpha^{\;\;A} X_A\equiv \nabla^\top_\alpha+i \left(E_\alpha^{\;\;\mu} L_{4\mu}+W_\alpha^{\mu\nu}L_{\mu\nu}\right),\;\; {K}_\alpha^{\;\;A} \equiv (E_\alpha^{\;\;\mu},W_\alpha^{\mu\nu}).
\end{equation}
The solution of the field equation (\ref{fer31}) is recalled in the following subsection, and then the tensor fields of different ranks are presented.

\subsection{Gauge vector field}

In terms of the dS plane wave, the solution of the equation (\ref{fer31}) can be written as \cite{gagarota}:
\begin{equation}
\mathcal{K}_\alpha^{\;\;A}(x)=\mathcal{E}_\alpha^{\;\;A}(x,\xi,Z^A)(x\cdot\xi)^\sigma,\;\;\; \sigma=-1,-2\, ;\;\; \xi\cdot \xi=0,
\end{equation}
where $Z_\alpha^A$ are ten arbitrary constant five-vectors and $\xi^\alpha=(\xi^0, \vec \xi, \xi^4)\in C^+$ is the null five-vector and $C^+$ is the positive cone, which is defined by:
\begin{equation} \label{cplas}
C^+=\left\lbrace\xi \in \mathrm{I\hspace{-.15em}R}^5 |\; \; \xi \cdot
\xi=0 , \;\;\xi^0>0\right\rbrace.
\end{equation}
In the following, for simplicity, we ignore the index $A$. The tensor field can be constructed by using these vector fields, so the polarization vector and its properties have briefly been recalled as follows \cite{gagarota}:
\begin{equation}
\mathcal{E}_\alpha(x,\xi,Z)=\left({Z}^\top_\alpha- \frac{Z\cdot
x}{x\cdot\xi }\,{\xi}^\top_\alpha\right)=\left({Z}_\alpha-\frac{Z\cdot
x}{x\cdot\xi}\,{\xi}_\alpha\right).
\end{equation}
By choosing $\xi\cdot Z=0$, the vector polarization satisfies the equations $\partial^\top \cdot \mathcal{E}=0,\; Q_0\; \mathcal{E}_\alpha=0$.

The simplest form of $Z_{\alpha}$ compatible with the Minkowski polarization vector in the flat
limit $\epsilon_{\mu}^{(\lambda)}(k)$ is achieved if the $Z^{(\lambda)}_\alpha$'s satisfy the following conditions \cite {gata}:
\begin{equation}\label{eq:pola1} Z^{(\lambda)}\cdot
Z^{(\lambda')}=-\delta^{\lambda\lambda'},\quad\sum_{\lambda} Z^{(\lambda)}_{\alpha}Z^{(\lambda)}_{\beta}
=-\eta_{\alpha\beta} \,,
\end{equation}
where $\lambda$'s are the polarizations of the vector field. A remarkable feature of using ambient space notation is that the properties of the dS polarization vector are very similar to the Minkowskian case:
\begin{eqnarray}\label{eq:pola2}
&&\sum_{\lambda}\mathcal{E}^{(\lambda)}_{\alpha}(x,\xi,Z)\,\mathcal{E}^{(\lambda)}_{\beta}(x,\xi,Z)=-\left(\theta_{\alpha\beta}-\frac{\xi^\top_{\alpha}
\xi^\top_{\beta}}{(x\cdot\xi)^2}\right)\equiv \Pi_{\alpha\beta}(x,\xi)\,,\nonumber\\
&&\mathcal{E}^{(\lambda)}(x,\xi,Z)\cdot\mathcal{E}^{(\lambda')}(x,\xi,Z)=Z^{(\lambda)}\cdot\mathcal{E}^{(\lambda')}(x,\xi,Z)=\mathcal {E}^{(\lambda)}(x,\xi,Z)\cdot\mathcal{E}^{(\lambda')}(x',\xi,Z)=-\delta^{\lambda\lambda'} \,.
\end{eqnarray}

The following identities which are necessary for the construction of the other tensor fields have been proved:
\b
\partial^\top_\beta \mathcal{E}_\alpha=-\xi_\alpha\frac{Z^\top_\beta}{x\cdot \xi}+\xi_\alpha\frac{x\cdot Z}{(x\cdot \xi)2}\xi^\top_\beta,
\;\;\; \xi^{\top\beta}\partial^\top_\beta \mathcal{E}_\alpha=0,
\e
\b
\mathcal{E}^\beta \partial^\top_\beta \mathcal{E}_\alpha=-\frac{Z\cdot Z}{x\cdot \xi}\xi_\alpha,\;\;\;\;\;
\partial^\top_\beta \mathcal{E}_\alpha \partial^{\top\beta }\mathcal{E}'_\gamma=\frac{Z\cdot Z'}{(x\cdot \xi)^2}\xi_\alpha\xi_\gamma,
\e
\b
Q_0 \theta_{\alpha\beta}=-8 x_\alpha x_\beta-2 \theta_{\alpha\beta}, \;\;\; Q_0\mathcal{E}_\alpha=0,\;\;\; Q_0(x\cdot \xi)^\sigma =-\sigma(\sigma+3)(x\cdot\xi)^\sigma.
\e

\subsection{Gauge rank-$2$ tensor field}

The spin-$2$ field can be constructed from the two vector gauge fields $E^\mu_\alpha $ from relation $H_{\alpha \beta}=E^\mu_\alpha E^\nu_\beta q_{\mu\nu}$ \cite{mama}, where $q_{\mu\nu}$ is a numerical constant, which can be chosen as $q_{\mu\nu}=\delta_{\mu\nu}$. Using the above gauge vector fields, one can construct the spin-$2$ rank-$2$ symmetric tensor field $H_{\alpha \beta}$ in the following form:
\begin{equation}\label{eq:sol}
H_{\alpha\beta}(x)=\mathcal {E}_{\alpha\beta}(x,\xi,Z)(x\cdot\xi)^{\sigma},\end{equation}
where the homogeneity degree $\sigma$ is an arbitrary constant that can be fixed by applying the physical conditions on the spin-$2$ field. Generalized polarization tensor is \cite{gagata}:
\begin{equation}\label{eq:tensor1}
\mathcal{E}_{\alpha\beta}^{(\lambda\lambda')}(x,\xi)= \frac{1}{2}\left[
\mathcal{S}\,\mathcal{E}_{\alpha}^{(\lambda)}(x,\xi)\; \mathcal{E}_{\beta}^{(\lambda')}(x,\xi)-\frac{2}{3}\, \delta^{\lambda\lambda'}
\sum_{(\rho)}\mathcal{E}^{(\rho)}_{\alpha}(x,\xi)\mathcal{E}^{(\rho)}_{\beta}(x,\xi) \right],
\end{equation}
where $\mathcal{S}$ denotes the symmetrisation indices and $\mathcal{E}^{(\lambda)}(x,\xi)=\mathcal{E}^{(\lambda)}(x,\xi,Z)$. One can easily check that the tensor polarization
(\ref{eq:tensor1}) satisfies the following properties (see appendix B):
\b \label{prsp2} \eta^{\alpha\beta}\mathcal{E}_{\alpha\beta}(x,\xi,Z)=0, \;\; \xi^\top \cdot \mathcal{E}_{\cdot\beta}(x,\xi,Z)=0, \e
\b \label{prsp3} \mathcal{E}_{\alpha\beta}=\mathcal{E}_{\beta\alpha},\;\;\xi^{\top\gamma} \partial^\top_\gamma \mathcal{E}_{\alpha\beta}=0,\;\; Q_0 \mathcal{E}_{\alpha\beta}=0.\e
It can be easily shown that this tensor field satisfies the following equation (see appendix B):
$$
Q_0H_{\alpha\beta}(x)=\left[Q_0 \mathcal{E}_{\alpha\beta}\right](x\cdot\xi)^{\sigma}-\sigma(\sigma+3)\mathcal{E}_{\alpha\beta}(x\cdot\xi)^{\sigma}-
2[\partial^\top_\gamma \mathcal{E}_{\alpha\beta}
]\xi^{\top\gamma}\sigma (x\cdot\xi)^{\sigma-1}$$
\b \label{prsp4} =-\sigma(\sigma+3)\mathcal{E}_{\alpha\beta}(x\cdot\xi)^{\sigma}. \e
This field transform as an UIR of the dS group $\Pi^\pm_{2,2}$ in the Dixmir notation \cite{dix}, if we have $Q_0H_{\alpha\beta}(x)=0$. Then we obtain $\sigma=0,-3$.

Barut and B{\"o}hm have shown that the value of the Casimir operator of the conformal group for the unitary irreducible representation of the spin-$2$ field must be equal to $9$ \cite{babo}. However, Binegar et al. proved that for a rank-$2$ symmetric tensor field, it becomes $8$, and for a mixed symmetry rank-$3$ tensor field, it can be fixed to $9$ \cite{bifrhe}. Although the spin-$2$ rank-$2$ tensor field transforms by the UIR of the dS group, it breaks the conformal transformation. To preserve the conformal transformation, the most straightforward choice at the moment is a spin-$2$ mixed symmetry rank-$3$ tensor field \cite{ta1403}. This field is constructed in the following subsection.

\subsection{Gauge rank-$3$ tensor field}

A spin-$2$ mixed symmetry rank-$3$ tensor field on the dS hyperboloid is defined as:
$$\mathcal{K}_{\alpha \beta\gamma}=-\mathcal{K}_{\alpha\gamma\beta}, \;\;\; \mathcal{K}_{\alpha \beta\gamma}+\mathcal{K}_{\gamma\alpha \beta}+\mathcal{K}_{\beta\gamma\alpha }=0, \;\;\; x \cdot \mathcal{K}=0\,. $$
By imposing the following subsidiary conditions on this tensor field:
\b \partial \cdot \mathcal{K}=0,\;\;\; \eta^{\alpha\beta}\mathcal{K}_{\alpha \beta\gamma}=0,\;\;\; Q_0\mathcal{K}_{\alpha \beta\gamma}=0,\e
it can be simultaneously transformed as a UIR of the dS group, and the conformal group \cite{ta1403,derotata}.

In terms of the rank-$2$ symmetric tensor field and the vector field, this field can be derived as:
\begin{equation}\label{eq12}
\mathcal{K}_{\alpha\beta\gamma}=\mathcal{E}_{\alpha\beta\gamma}(x,\xi,Z,Y)(x \cdot \xi)^\sigma\,,
\end{equation}
where
\begin{equation}\label{eq13}
\mathcal{E}_{\alpha\beta\gamma}(x,\xi,Z,Y)=\mathcal{E}_{\alpha\beta}(x,\xi,Z)\mathcal{E}_\gamma(x,\xi,Y)-\mathcal{E}_{\alpha\gamma}(x,\xi,Z)\mathcal{E}_\beta(x,\xi,Y),
\end{equation}
$Z_\alpha$ and $ Y_\beta$ are two constant arbitrary vector fields. By applying the condition $Z\cdot Y=0$, the following identities can be easily proved:
\begin{equation}\label{eq14}
\mathcal{E}_{\alpha\beta}(x,\xi,Z)\mathcal{E}^\beta(x,\xi,Y)=0,\;\;\;\xi^{\top\delta} \partial^\top_\delta \mathcal{E}_{\alpha\beta\gamma}(x,\xi,Z,Y)=0,
\end{equation}
\begin{equation}\label{eq15}
Q_0 \mathcal{E}_{\alpha\beta\gamma}(x,\xi,Z,Y)=0.
\end{equation}
Then, the tensor field (\ref{eq12}) satisfies the following field equation:
$$ Q_0\K_{\alpha\beta\gamma}(x)=(Q_0 \mathcal{E}_{\alpha\beta\gamma}(x,\xi,Z,Y))(x \cdot \xi)^\sigma+\mathcal{E}_{\alpha\beta\gamma}(x,\xi,Z,Y)Q_0((x \cdot \xi)^\sigma)$$ \b =-\sigma(\sigma+3)\mathcal{E}_{\alpha\beta\gamma}(x,\xi,Z,Y)(x\cdot\xi)^{\sigma}.\e
The condition, $Q_0\K_{\alpha\beta\gamma}(x)=0$, is necessary for the conformal transformation and results to $\sigma=0,-3$ \cite{ta1403,tatafa,fatata}. The conformal invariant field equations for the massless fields in dS ambient space formalism were obtained from the group theoretical point of view \cite{derotata,tatafa,taro12,fatata}. The spin-$2$ rank-$2$ symmetric tensor field and its conformal field equation have been obtained \cite{derotata}. A mixed symmetry rank-$3$ gauge field has been considered in our work. Moreover, their related tensor polarizations have been constructed. In \cite{tatafa}, the group's theoretical point of view has defined the relation between rank-$2$ symmetric and mixed symmetry rank-$3$ tensor fields. This paper introduces a rank-$3$ tensor polarization by using the gauge theory.

\section{conclusion}

The gauge principle governs the interactions between the elementary systems in the universe. These interactions are formulated through the gauge-covariant derivative. They are defined as the quantities that preserve the gauge-invariant transformation of the Lagrangian. In this paper, the spin-$2$ gauge field has been presented from the gauge vector fields of the dS group in the line of the Yang-Mills gauge theory. It is important to note that the rank-$2$ symmetric tensor field is written regarding a combination of the vector gauge fields. Unfortunately, the usual rank-$2$ symmetric tensor field breaks the conformal transformation, although it can lead to the super-gravity gauge theory in dS universe \cite{taksgt}. To preserve the conformal invariant, a mixed symmetry rank-$3$ gauge field must be used. The mixed symmetry rank-$3$ tensor field can be written in rank-$2$ symmetric tensor and vector gauge fields. These results help us to formulate the super-conformal gravity in the dS ambient space formalism \cite{rata}.


\vspace{0.5cm} \noindent {\bf{Acknowlegements}}: We are grateful
to Maryam Amiri for her interest in this work.
\appendix

\setcounter{equation}{0}
\section{Field equation}

The field equation (\ref{casira2}) is derived in this appendix. Substituting equation (\ref{cuso14}) in equation (\ref{gias2r31gg}), and doing some simplifications, one can obtain:
$$ S[{K}]=\int d\mu(x)\left(\nabla^\top_\alpha{K}_\beta^{\;\;A}q_{AB}\nabla^{\top\alpha} {K}^{\beta{\;\;B}}-
\nabla^\top_\alpha{K}_\beta^{\;\;A}q_{AB}\nabla^{\top\beta }{K}^{\alpha{\;\;B}}
+\nabla^\top_\alpha{K}_\beta^{\;\;A}q_{AB}{K}^{\beta{\;\;E}}{K}^{\alpha{\;\;F}}f_{EF}^{\;\;\;\;B}\right.$$
$$\left.-\nabla^\top_\beta{K}_\alpha^{\;\;A}q_{AB}\nabla^{\top\alpha }{K}^{\beta{\;\;B}}+\nabla^\top_\beta{K}_\alpha^{\;\;A}q_{AB}\nabla^{\top\beta }{K}^{\alpha{\;\;B}}-\nabla^\top_\beta{K}_\alpha^{\;\;A}q_{AB}{K}^{\beta{\;\;E}}{K}^{\alpha{\;\;F}}f_{EF}^{\;\;\;\;B}\right.$$
\b \label{gias2r31}
\left. +{K}_\beta^{\;\;C}{K}_\alpha^{\;\;D}f_{CD}^{\;\;\;\;A} q_{AB} \nabla^{\top\alpha }{K}^{\beta{\;\;B}}-{K}_\beta^{\;\;C}{K}_\alpha^{\;\;D}f_{CD}^{\;\;\;\;A} q_{AB}\nabla^{\top\beta} {K}^{\alpha{\;\;B}}
+ {K}_\beta^{\;\;C}{K}_\alpha^{\;\;D}f_{CD}^{\;\;\;\;A} q_{AB}{K}^{\beta{\;\;E}}{K}^{\alpha{\;\;F}}f_{EF}^{\;\;\;\;B}\right)
.\e
The field equation is obtained from the Euler-Lagrange equation :
\b \label{Euler}\nabla^\top_{\alpha'}\left(\frac{\delta\mathcal{L}}{\delta(\nabla^\top_{\alpha'}\mathcal{K}_{\beta'}^{\;\;A'})}\right)-\frac{\delta\mathcal{L}}{\delta\mathcal{K}_{\beta'}^{\;\;A'}}=0.\e
The first term, after some simplifications, is:
$$\frac{\delta\mathcal{L}}{\delta(\nabla^\top_{\alpha'}{K}_{\beta'}^{\;\;A'})}=2\left[\nabla^{\top\alpha'} {K}^{\beta'B}-\nabla^{\top\beta'} {K}^{\alpha'B} \right]q_{A'B}+2\left[\nabla^{\top\alpha'} {K}^{\beta'A}-\nabla^{\top\beta'} {K}^{\alpha'A}\right]q_{AA'}$$
\b \label{gerad1}
+\left[{K}^{\beta' E}{K}^{\alpha' F}-{K}^{\alpha' E}{K}^{\beta' F}\right]f_{EF}^{\;\;\;\;B}q_{A'B}+\left[{K}^{\beta' C}{K}^{\alpha' D}-{K}^{\alpha' C}{K}^{\beta' D}\right]f_{CD}^{\;\;\;\;A}q_{AA'},\e
and the second term in (\ref{Euler}) can be calculated as:
$$\frac{\delta\mathcal{L}}{\delta\mathcal{K}_{\beta'}^{\;\;A'}}= \left[\nabla^\top_\alpha{K}^{\beta'A}-\nabla^{\top\beta'}{K}_\alpha^{\;\;A}\right]{K}^{\alpha F}f_{A'F}^{\;\;\;\;B}q_{AB}+\left[\nabla^{\top\beta'}{K}_\beta^{\;\;A}-\nabla^\top_\beta{K}^{\beta' A}\right]{K}^{\beta E}f_{EA'}^{\;\;\;\;B}q_{AB}$$
$$+\left[\nabla^{\top\alpha}{K}_{\beta'}^{\;\;B}-\nabla^{\top\beta'}{K}^{\alpha B}\right]{K}_\alpha^{\;\;D}f_{A'D}^{\;\;\;\;A}q_{AB}+\left[\nabla^{\top\beta'}{K}^{\beta B}-
\nabla^{\top\beta}{K}^{\beta' B}\right]{K}_\beta^{\;\;C}f_{CA'}^{\;\;\;\;A}q_{AB}$$
$$
+\left[{K}_\alpha^{\;\;D}f_{A'D}^{\;\;\;\;A}{K}^{\beta' E}{K}^{\alpha F}+{K}_\beta^{\;\;C}f_{CA'}^{\;\;\;\;A}{K}^{\beta E}{K}^{\beta' F}\right]f_{EF}^{\;\;\;\;B}q_{AB}$$
\b \label{Euler1}
+\left[{K}^{\beta' C}{K}_\alpha^{\;\;D}{K}^{\alpha F}f_{A'F}^{\;\;\;\;B}+{K}_\beta^{\;\;C}{K}^{\beta' D}{K}^{\beta E}f_{EA'}^{\;\;\;\;B}\right]f_{CD}^{\;\;\;\;A}q_{AB}.\e

By substitution of (\ref{Euler1}) and (\ref{gerad1}) into (\ref{Euler}), the field equation can be derived as:
$$ 2\nabla^\top_{\alpha'}\left[\nabla^{\top\alpha'} {K}^{\beta'B}-\nabla^{\top\beta'} {K}^{\alpha'B}\right]q_{A'B}+2\nabla^\top_{\alpha'}\left[\nabla^{\top\alpha'} {K}^{\beta'A}-\nabla^{\top\beta'} {K}^{\alpha'A}\right]q_{AA'}$$ $$+\nabla^\top_{\alpha'}\left[{K}^{\beta' E}{K}^{\alpha' F}-{K}^{\alpha' E}{K}^{\beta' F}\right]f_{EF}^{\;\;\;\;B}q_{A'B}+\nabla^\top_{\alpha'}\left[{K}^{\beta' C}{K}^{\alpha' D}-{K}^{\alpha' C}{K}^{\beta' D}\right]f_{CD}^{\;\;\;\;A}q_{AA'}$$
$$-\left[\nabla^\top_\alpha{K}^{\beta'A}-\nabla^{\top\beta'}{K}_\alpha^{\;\;A}\right]{K}^{\alpha F}f_{A'F}^{\;\;\;\;B}q_{AB}-\left[\nabla^{\top\beta'}{K}_\beta^{\;\;A}-\nabla^\top_\beta{K}^{\beta' A}\right]{K}^{\beta E}f_{EA'}^{\;\;\;\;B}q_{AB}$$
$$-\left[\nabla^{\top\alpha}{K}_{\beta'}^{\;\;B}-\nabla^{\top\beta'}{K}^{\alpha B}\right]{K}_\alpha^{\;\;D}f_{A'D}^{\;\;\;\;A}q_{AB}-\left[\nabla^{\top\beta'}{K}^{\beta B}- \nabla^{\top\beta}{K}^{\beta' B}\right]{K}_\beta^{\;\;C}f_{CA'}^{\;\;\;\;A}q_{AB}$$
$$ -\left[{K}_\alpha^{\;\;D}f_{A'D}^{\;\;\;\;A}{K}^{\beta' E}{K}^{\alpha F}+{K}_\beta^{\;\;C}f_{CA'}^{\;\;\;\;A}{K}^{\beta E}{K}^{\beta' F}\right]f_{EF}^{\;\;\;\;B}q_{AB} $$
\b \label{gerad2} -\left[{K}^{\beta' C}{K}_\alpha^{\;\;D}{K}^{\alpha F}f_{A'F}^{\;\;\;\;B}+{K}_\beta^{\;\;C}{K}^{\beta' D}{K}^{\beta E}f_{EA'}^{\;\;\;\;B}\right]f_{CD}^{\;\;\;\;A}q_{AB}=0.\e
The linear field equation or free field equation is:
$$2\nabla^\top_{\alpha'}\left[\nabla^{\top\alpha'} {K}^{\beta'B}-\nabla^{\top\beta'} {K}^{\alpha'B} \right]q_{A'B}+2\nabla^\top_{\alpha'}\left[\nabla^{\top\alpha'} {K}^{\beta'A}-\nabla^{\top\beta'} {K}^{\alpha'A}\right]q_{AA'}=0.$$
Since the two terms in the above relation are equivalent, the linear field equation becomes:
\b \label{gerad2} \nabla^\top_{\alpha'}\nabla^{\top\alpha'} {K}^{\beta'A}-\nabla^\top_{\alpha'}\nabla^{\top\beta'} {K}^{\alpha'A}=0.\e
Using the following commutation relation \cite{ta1403}:
\begin{equation} \label{trons} \left[\nabla^{\top}_{\alpha'}, \nabla^{\top\beta'}\right]{K}^{\alpha'A}\equiv \left(R^{dS}\right)_{\lambda\;\;\alpha'}^{\;\;\alpha'\;\;\beta'}{K}^{\lambda A} =\left(\theta_{\lambda\alpha'}\theta^{\alpha'\beta'} {K}^{\lambda A}- \theta_{\alpha'}^{\alpha'}\theta^{\beta'}_\lambda {K}^{\lambda A}\right)
=-3 {K}^{\beta'A},
\end{equation}
the equation (\ref{gerad2}) can be rewritten to the following form:
\begin{equation} \label{trons1}
\nabla^\top_{\alpha'} \nabla^{\top\alpha'} {K}^{\beta'A}-\nabla^{\top\beta'}\nabla^\top_{\alpha'} {K}^{\alpha'A}+3 {K}^{\beta'A}=0.
\end{equation}
The first two terms in (\ref{trons1}) by using relation (\ref{dscdrt}) can be rewrite as:
\begin{equation} \label{trons2}
\nabla^\top_{\alpha'} \nabla^{\top\alpha'} {K}^{\beta'A}=-Q_0 {K}^{\beta'A}- {K}^{\beta'A}-x^{\beta'} \partial^\top \cdot{K}^{\;A},
\end{equation}
\begin{equation} \label{trons3}
\nabla^{\top\beta'}\nabla^\top_{\alpha'} {K}^{\alpha'A}= \partial^{\top\beta'}\partial^\top \cdot{K}^{\;A}.
\end{equation}
By substitution of (\ref{trons2}) and (\ref{trons3}) in (\ref{trons1}), the field equation (\ref{casira2}) is proved.

\setcounter{equation}{0}
\section{Proof of some relations}

In this appendix the equations (\ref{prsp3}) and (\ref{prsp4}) are proved. One can calculate the derivative of (\ref{eq:tensor1}) as follows:
$$\partial^{\top\gamma} \mathcal{E}_{\alpha\beta}=\frac{1}{2}\left[-\xi_\alpha\frac{{Z}^{\top(\lambda)}_\gamma{Z}^{(\lambda')}_\beta}{(x\cdot \xi)}-\xi_\beta\frac{{Z}^{\top(\lambda)}_\gamma{Z}^{(\lambda')}_\alpha}{(x\cdot \xi)}-\xi_\alpha\frac{{Z}^{(\lambda)}_\beta{Z}^{\top(\lambda')}_\gamma}{(x\cdot \xi)}-\xi_\beta\frac{{Z}^{(\lambda)}_\alpha{Z}^{\top(\lambda')}_\gamma}{(x\cdot \xi)}+\xi_\alpha\xi^\top_\gamma\frac{{Z}^{(\lambda')}_\beta(x\cdot {Z}^{(\lambda)})}{(x\cdot \xi)^2}\right.$$
$$ \left.+\xi_\beta\xi^\top_\gamma\frac{{Z}^{(\lambda')}_\alpha(x\cdot {Z}^{(\lambda)})}{(x\cdot \xi)^2}+\xi_\alpha\xi^\top_\gamma\frac{{Z}^{(\lambda)}_\beta(x\cdot {Z}^{(\lambda')})}{(x\cdot \xi)^2}+\xi_\beta\xi^\top_\gamma\frac{{Z}^{(\lambda)}_\beta(x\cdot {Z}^{(\lambda')})}{(x\cdot \xi)^2}\right]+\xi_\alpha\xi_\beta\frac{{Z}^{\top(\lambda)}_\gamma(x\cdot {Z}^{(\lambda')})}{(x\cdot \xi)^2}$$
$$ +\xi_\alpha\xi_\beta\frac{{Z}^{\top(\lambda')}_\gamma(x\cdot {Z}^{(\lambda)})}{(x\cdot \xi)^2}-2\xi_\alpha\xi_\beta\xi^\top_\gamma\frac{(x\cdot {Z}^{(\lambda)})(x\cdot {Z}^{(\lambda')})}{(x\cdot \xi)^3}-\frac{1}{3}\delta^{(\lambda\lambda')}\sum_{\rho}\left[-\xi_\alpha\frac{{Z}^{(\rho)}_\beta {Z}^{\top(\rho)}_\gamma}{(x\cdot \xi)}-\xi_\beta\frac{{Z}^{(\rho)}_\alpha{Z}^{\top(\rho)}_\gamma}{(x\cdot \xi)}\right.$$
\begin{equation} \label{polar5}
\left. +2\xi_\alpha\xi_\beta\frac{{Z}^{\top(\rho)}_\gamma(x\cdot {Z}^{(\rho)})}{(x\cdot \xi)^2}+\xi_\alpha\xi^\top_\gamma\frac{{Z}^{(\rho)}_\beta(x\cdot {Z}^{(\rho)})}{(x\cdot \xi)^2}+\xi_\beta\xi^\top_\gamma\frac{{Z}^{(\rho)}_\alpha(x\cdot {Z}^{(\rho)})}{(x\cdot \xi)^2}-2\xi_\alpha\xi_\beta\xi^\top_\gamma\frac{(x\cdot {Z}^{(\rho)})^2}{(x\cdot \xi)^3}\right].
\end{equation}
Now using this relation, one can obtain the equations (\ref{prsp3}):
$$\xi^\top_{\gamma}\partial^{\top\gamma} \mathcal{E}_{\alpha\beta}=\frac{1}{2}\left[-\xi_\alpha\frac{(x\cdot {Z}^{(\lambda)})(x\cdot \xi){Z}^{(\lambda')}_\beta}{(x\cdot \xi)}-\xi_\beta\frac{(x\cdot {Z}^{(\lambda)})(x\cdot \xi){Z}^{(\lambda')}_\alpha}{(x\cdot \xi)}-\xi_\alpha\frac{{Z}^{(\lambda)}_\beta(x\cdot {Z}^{(\lambda')})(x\cdot \xi)}{(x\cdot \xi)}\right.$$
$$ \left.-\xi_\beta\frac{{Z}^{(\lambda)}_\alpha(x\cdot {Z}^{(\lambda')})(x\cdot \xi)}{(x\cdot \xi)}+\xi_\alpha\frac{{Z}^{(\lambda')}_\beta(x\cdot {Z}^{(\lambda)})(x\cdot \xi)^2}{(x\cdot \xi)^2}+\xi_\beta\frac{{Z}^{(\lambda')}_\alpha(x\cdot {Z}^{(\lambda)})(x\cdot \xi)^2}{(x\cdot \xi)^2}+\xi_\alpha\frac{{Z}^{(\lambda)}_\beta(x\cdot {Z}^{(\lambda')})(x\cdot \xi)^2}{(x\cdot \xi)^2}\right.$$
$$\left.+\xi_\beta\frac{{Z}^{(\lambda)}_\beta(x\cdot {Z}^{(\lambda')})(x\cdot \xi)^2}{(x\cdot \xi)^2}\right]+\xi_\alpha\xi_\beta\frac{(x\cdot {Z}^{(\lambda)})(x\cdot \xi)(x\cdot {Z}^{(\lambda')})}{(x\cdot \xi)^2}+\xi_\alpha\xi_\beta\frac{(x\cdot {Z}^{(\lambda')})(x\cdot \xi)(x\cdot {Z}^{(\lambda)})}{(x\cdot \xi)^2}$$
$$-2\xi_\alpha\xi_\beta\frac{(x\cdot {Z}^{(\lambda)})(x\cdot {Z}^{(\lambda')})(x\cdot \xi)^2}{(x\cdot \xi)^3}-\frac{1}{3}\delta^{(\lambda\lambda')}\sum_{\rho}\left[-\xi_\alpha\frac{{Z}^{(\rho)}_\beta(x\cdot {Z}^{(\rho)})(x\cdot \xi)}{(x\cdot \xi)}-\xi_\beta\frac{{Z}^{(\rho)}_\alpha(x\cdot {Z}^{(\rho)})(x\cdot \xi)}{(x\cdot \xi)}\right.$$
\begin{equation} \label{polar5}
\left.+2\xi_\alpha\xi_\beta\frac{(x\cdot {Z}^{(\rho)})^2(x\cdot \xi)}{(x\cdot \xi)^2}+\xi_\alpha\frac{{Z}^{(\rho)}_\beta(x\cdot \xi)^2(x\cdot {Z}^{(\rho)})}{(x\cdot \xi)^2}+\xi_\beta\frac{{Z}^{(\rho)}_\alpha(x\cdot \xi)^2(x\cdot {Z}^{(\rho)})}{(x\cdot \xi)^2}-2\xi_\alpha\xi_\beta\frac{(x\cdot {Z}^{(\rho)})^2(x\cdot \xi)^2}{(x\cdot \xi)^3}\right]=0,
\end{equation}
and
$$Q_0 \mathcal{E}_{\alpha\beta}=-\partial^{\top}_\gamma\partial^{\top\gamma} \mathcal{E}_{\alpha\beta}=-\frac{1}{2}\left[-3\xi_\alpha\frac{{Z}^{(\lambda')}_\beta(x\cdot {Z}^{(\lambda)})}{(x\cdot \xi)}-3\xi_\beta\frac{{Z}^{(\lambda')}_\alpha(x\cdot {Z}^{(\lambda)})}{(x\cdot \xi)}-3\xi_\alpha\frac{{Z}^{(\lambda)}_\beta(x\cdot {Z}^{(\lambda')})}{(x\cdot \xi)}-3\xi_\beta\frac{{Z}^{(\lambda)}_\alpha(x\cdot {Z}^{(\lambda')})}{(x\cdot \xi)}\right.$$
$$\left.+3\xi_\alpha\frac{{Z}^{(\lambda')}_\beta(x\cdot {Z}^{(\lambda)})}{(x\cdot \xi)}+3\xi_\beta\frac{{Z}^{(\lambda')}_\alpha(x\cdot {Z}^{(\lambda)})}{(x\cdot \xi)}+3\xi_\alpha\frac{{Z}^{(\lambda)}_\beta(x\cdot {Z}^{(\lambda')})}{(x\cdot \xi)}+3\xi_\beta\frac{{Z}^{(\lambda)}_\alpha(x\cdot {Z}^{(\lambda')})}{(x\cdot \xi)}\right]$$
$$+6\xi_\alpha\xi_\beta\frac{(x\cdot {Z}^{(\lambda)})(x\cdot {Z}^{(\lambda')})}{(x\cdot \xi)^2}+2\xi_\alpha\xi_\beta\frac{({Z}^{(\lambda)}\cdot{Z}^{(\lambda')})}{(x\cdot \xi)^2}-6\xi_\alpha\xi_\beta\frac{(x\cdot {Z}^{(\lambda)})(x\cdot {Z}^{(\lambda')})}{(x\cdot \xi)^2}-\frac{1}{3}\delta^{(\lambda\lambda')}\sum_{(\rho)}$$
$$\left[-3\xi_\alpha\frac{{Z}^{(\rho)}_\beta(x\cdot {Z}^{(\rho)})}{(x\cdot \xi)}-3\xi_\beta\frac{{Z}^{(\rho)}_\alpha(x\cdot {Z}^{(\rho)})}{(x\cdot \xi)}+6\xi_\alpha\xi_\beta\frac{(x\cdot {Z}^{(\rho)})^2}{(x\cdot \xi)^2}+2\xi_\alpha\xi_\beta\frac{{Z}^{(\rho)}\cdot{Z}^{(\rho)}}{(x\cdot \xi)^2}+3\xi_\alpha\frac{{Z}^{(\rho)}_\beta(x\cdot {Z}^{(\rho)})}{(x\cdot \xi)}\right.$$
\begin{equation} \label{polar6}
\left.+3\xi_\beta\frac{{Z}^{(\rho)}_\alpha(x\cdot {Z}^{(\rho)})}{(x\cdot \xi)}-6\xi_\alpha\xi_\beta\frac{(x\cdot {Z}^{(\rho)})^2}{(x\cdot \xi)^2}\right]=2\xi_\alpha\xi_\beta\frac{({Z}^{(\lambda)}\cdot{Z}^{(\lambda')})}{(x\cdot \xi)^2}-\frac{1}{3}\delta^{(\lambda\lambda')}\sum_{(\rho)}\left[2\xi_\alpha\xi_\beta\frac{{Z}^{(\rho)}\cdot{Z}^{(\rho)}}{(x\cdot \xi)^2}\right]=0.
\end{equation}

Then the equation (\ref{prsp4}) can be easily calculated:
$$Q_0H_{\alpha\beta}(x)=Q_0 \left[\mathcal{E}_{\alpha\beta}(x,\xi,Z)(x\cdot\xi)^{\sigma}\right]$$
$$=-\partial^{\top\gamma}\partial^{\top}_\gamma\left[\mathcal{E}_{\alpha\beta}(x,\xi,Z)(x\cdot\xi)^{\sigma}\right]=-\partial^{\top\gamma}\left[(\partial^{\top}_\gamma\mathcal{E}_{\alpha\beta})(x\cdot \xi)^{\sigma}+\mathcal{E}_{\alpha\beta}\partial^{\top}_\gamma(x\cdot \xi)^{\sigma}\right]$$
$$=Q_0\mathcal{E}_{\alpha\beta}-2{\sigma}(\partial^{\top}_\gamma\mathcal{E}_{\alpha\beta})\xi^{\top\gamma}(x\cdot \xi)^{{\sigma}-1}-{\sigma}({\sigma}-1)\mathcal{E}_{\alpha\beta}\xi^{\top\gamma}(x\cdot \xi)^{{\sigma}-2}\xi^{\top}_\gamma-4{\sigma}\mathcal{E}_{\alpha\beta}(x\cdot \xi)^{{\sigma}-1}(x\cdot \xi)$$
\begin{equation} \label{polar7}
=Q_0\mathcal{E}_{\alpha\beta}-{\sigma}({\sigma}+3)\mathcal{E}_{\alpha\beta}(x\cdot \xi)^{\sigma}-2{\sigma}\left[\partial^{\top}_{\gamma}\mathcal{E}_{\alpha\beta}\right]\xi^{\top\gamma}(x\cdot\xi)^{{\sigma}-1}=-{\sigma}({\sigma}+3)\mathcal{E}_{\alpha\beta}(x\cdot \xi)^{\sigma}.
\end{equation}


\begin{thebibliography}{99}

\bibitem{acce} A.G. Riess et al., Astron. J. \textbf{116} 1009 (1998) \textit{Observational evidence from supernovae for an accelerating universe and a cosmological constant}, [\href{https://arxiv.org/abs/astro-ph/9805201}{arXiv:astro-ph/9805201}].

\bibitem{perme} S. Perlmutter et al., Astrophys J. \textbf{517} 565 (1999) \textit{Measurement of $\Omega$ and $\lambda$ from 42 high-redshift supernova}, [\href{https://arxiv.org/abs/astro-ph/9812133}{arXiv:astro-ph/9812133}].

\bibitem{hebb} J.P. Henry, U.G. Bohringer., Sci. Am. \textbf{279} 52 (1998) \textit{The evolution of galaxy clusters}.

\bibitem{hem} J.P. Henry, Astrophys. J. \textbf{534} 565 (2000) \textit{Measuring cosmological parameters from the evolution of cluster X-ray temperatures}.

\bibitem{berna} P. de Bernardis et al., Nature \textbf{404} 955 (2000) \textit{A flat universe from high-resolution maps of the cosmic microwave background radiation}, [\href{https://arxiv.org/abs/astro-ph/0004404}{arXiv:astro-ph/0004404}].

\bibitem{parde} P.A.R. Ade et al., Phys. Rev. Lett. \textbf{112} 241101 (2014) \textit{Detection of B-Mode polarization at degree angular scales by BICEP 2}.

\bibitem{ta1403} M.V. Takook, \textit{Quantum Field Theory in de Sitter Universe: Ambient Space Formalism}, [\href{https://arxiv.org/abs/1403.1204v3}{arXiv:gr-qc/1403.1204v3}].

\bibitem{ta2020} M.V. Takook, Int. J. Theor. Phys. {\bf 59}, 2540 (2020) {\it Conceptual and Technical Challenges
of Quantum Gravity}, [\href{https://arxiv.org/abs/2208.08098}{ arXiv:2208.08098}].

\bibitem{fr84} C. Fronsdal, Phys. Rev. D. \textbf{30} 2081 (1984) \textit{Ghost-free, nonlinear, spin-two, conformal gauge theory}.

\bibitem{SRI84} S. Deser and Rafael I. Nepomechie, Annals Of Physics. \textbf{154} 396 (1984) \textit{Gauge Invariance versus Masslessness in de Sitter Spaces}.

\bibitem{babo} A.O. Barut and A. B\"ohm, J. Math. Phys. \textbf{11} 2938 (1970) \textit{Reduction of a class of $O(4,2)$ representations with respect to $SO(4,1)$ and $SO(3,2)$}.

\bibitem{bifrhe} B. Binegar, C. Fronsdal, W. Heidenreich, Phys. Rev. D. \textbf{27} 2249 (1983) \textit{Linear conformal quantum gravity}.

\bibitem{nieto} J.A. Nieto, Phys. Rev. D. \textbf{50} R3583 (1994) \textit{Gauge theory of the de Sitter group and the Ashtekar formulation}.

\bibitem{mama} S.W. MacDowell, F. Mansouri, Phys. Rev. Lett. \textbf{38} 739 (1977) \textit{Unified geometric theory of gravity and supergravity}.

\bibitem{tak} B. Takahashi, Bull. Soc. Math. France \textbf{91} 289 (1963) \textit{Sur les repr\'esentations unitaires des groupes de Lorentz g\'en\'eralis\'es}.

\bibitem{tho} L.H. Thomas, Ann. of Math. \textbf{42} 113 (1941) \textit{On unitary representations of the group of de Sitter space}.

\bibitem{frhe} C. Fronsdal, W.F. Heidenreich, J. Math. Phys. \textbf{ 28} 215 (1978) \textit{Linear de Sitter gravity}.

\bibitem{gaha} J.P. Gazeau, M. Hans, J. Math. Phys. \textbf { 29} 2533 (1988) \textit{Integral-spin fields on $(3+2)$-de Sitter space}.

\bibitem{van} P. van Nieuwenhuizen, Phys. Rep. \textbf{68} 189 (1981) \textit{Supergravity}.

\bibitem{frts} E.S. Fradkin, A.A. Tseytlin, Phys. Rep. \textbf{119} 233 (1985) \textit{Conformal supergravity}.

\bibitem{gagarota}T. Garidi, J.P. Gazeau, S. Rouhani and M.V. Takook, J. Math. Phys. \textbf{49} 032501 (2008) \textit{ Massless vector field in de Sitter universe}, [\href{https://arxiv.org/gr-qc/0608004}{arXiv:gr-qc/0608004}].

\bibitem{gata} J.P. Gazeau, M.V. Takook, J. Math. Phys. \textbf{41} 5920 (2000) \textit{Massive vector field in de Sitter space}, [\href{https://arxiv.org/gr-qc/9912080}{arXiv:gr-qc/9912080}]; T. Garidi, J.P. Gazeau, M.V. Takook, ibid \textbf{43} 6379 (2002) \textit{Comment on: Massive vector field in de Sitter space}.

\bibitem{gagata} T. Garidi, J.P. Gazeau, S. Rouhani and M.V. Takook, J. Math. Phys. \textbf{44} 3838 (2003) \textit{Massive spin-2 field in de Sitter space}, [\href{https://arxiv.org/hep-th/0302022}{arXiv:gr-qc/0302022}].

\bibitem{dix} J. Dixmier, Bull. Soc. Math. France {\bf 89} 9 (1961) {\it Repr\'esentation int\'egrables du group de de Sitter}.

\bibitem{fatata} N. Fatahi, M.V. Takook, M.R. Tanhayi, Eur. Phys. J. C \textbf{74} 3111 (2014) \textit{Conformally covariant vector–spinor field in de Sitter space}, [\href{https://arxiv.org/1405.7535}{arXiv:1405.7535}]

\bibitem{derotata} M. Dehghani, S. Rouhani, M.V. Takook and M.R. Tanhay., Phys. Rev. D. \textbf{77} 064028 (2008) \textit{Conformally invariant massless spin-2 field in the de Sitter universe}, [\href{https://arxiv.org/0805.2227}{arXiv:0805.2227}].

\bibitem{tatafa} M.V. Takook, M. R. Tanhayi and S. Fatemi, J. Math. Phys. \textbf{51} 032503 (2010) \textit{Conformal linear gravity in de sitter space}, [\href{https://arxiv.org/0903.5249}{arXiv:0903.5249}].

\bibitem{taro12} M. Enayati, S. Rouhani, M.V. Takook, Int. J. Theor. Phys. \textbf{56} 1068 (2017) \textit{Quantum Linear Gravity in de Sitter Universe On Gupta-Bleuler vacuum state}, [\href{https://arxiv.org/abs/1208.5562}{arXiv:1208.5562}].

\bibitem{taksgt} M.V. Takook, \textit{de Sitter super-gravity in ambient space formalism}, [\href{https://arxiv.org/1712.09735v1}{arXiv:gr-qc/1712.09735v1}].

\bibitem{rata} R. Raziani, M.V. Takook, \textit{de Sitter super-conformal gravity in ambient space formalism}, in preparation.

\end{thebibliography}
\end{document}